\documentstyle[aps,prl,epsf, preprint]{revtex}

\begin{document}
\preprint{\vtop{\hbox{RU03-02-B}}}

\title{Color Superconductivity in a Dense Quark Matter}

\author{Hai-cang Ren
\footnote{E-mail: ren@summit.rockefeller.edu}}

\address{Physics Department, The Rockefeller University
1230 York Avenue, New York, NY 10021-6399}

\maketitle

\begin{abstract}
The color superconductivity of a dense quark matter is reviewed with emphasis 
on the long range nature of the pairing force and the multiplicity of the order parameter. 
The former gives rise to a non BCS behavior of the superconducting 
energy scale and the latter modifies the critical value of the Ginzburg-Landau
parameter that separates the superconductivity of type I and that of type II.
\end{abstract}


\ifpreprintsty\else
\fi

\section{Introduction}
\label{sec:Intro}
In this lecture, I would like to review our works for the past years on the 
color superconductivity in a dense quark matter. While it is an old subject dated 
back to 1970's ~\cite{BB} ~\cite{SF} ~\cite{DBAL}, the revived interests on the color 
superconductivity since 1998 ~\cite{ARW1} ~\cite{RSSV} ~\cite{ARW} are 
largely promoted by the experimental efforts of exploring the phase diagram of 
the strong interaction through RHIC or inside neutron stars. While lattice 
simulations have been successful at nonzero temperature, it becomes prohibitively difficult 
when the chemical potential becomes nonzero. Analytical technique 
remains the only means especially at large chemical potential, where an ideal 
gas of quarks and gluons presents the leading order approximation. The important progresses
that have been made recently include the discovery of the color-flavor locked structure in 
the super phase ~\cite{ARW} and the first-principle determination of the energy gap and the 
transition temperature ~\cite{DTS} ~\cite{SW} ~\cite{RPDR2} ~\cite{HMSW} 
~\cite{BLR1} ~\cite{BLR2} ~\cite{WQDR}. The entire subject
has been reviewed by several authors ~\cite{Rev}.
 
Among many novel characters of the 
color superconductivity, I would like to take this opportunity to highlight two of them which 
differ remarkably from the non-relativistic 
superconductivity in a metal. The first is the long range nature of the pairing 
force, which is responsible to the non BCS exponent of the weak coupling formula of 
the gap energy or the transition temperature and 
the suppression of the pre-exponential factor. The second is the multiplicity 
of the order parameters, which leads to a rich variety of inhomogeneous condensates  
and a different value of the critical Ginzburg-Landau 
parameter that separates the superconductivity of type I from that of type II. 
Many technical details have been suppressed and this lecture will serve a tour guide 
to our papers on the subject and the related works by others. 

\section{The Long Range Pairing Force.}
\label{sec:girep}
Let us consider a quark matter of ultra high baryon density such that the 
corresponding chemical potential is well above $\Lambda_{\rm QCD}$ to warrant 
a perturbative treatment because of the asymptotic freedom. While this density 
may not be attainable in a realistic quark matter, say the core of 
a neutron star, the approximation there is systematically under control. The 
Lagrangian density of QCD at a nonzero chemical potential $\mu$ in the chiral limit reads:
\begin{equation}
{\cal L}=-{1\over 2}{\rm tr}F_{\mu\nu}^lF_{\mu\nu}^l
-\bar\psi({\partial\over\partial x_\mu}-igA_\mu)\psi+\mu\bar\psi\gamma_4\psi+\hbox{renormalization 
counter terms},
\label{eq:lagrange}
\end{equation}
where
$A_\mu=A_\mu^lT^l$ and $F_{\mu\nu}={\partial A_\nu\over\partial x_\mu}
-{\partial A_\mu\over\partial x_\nu}-ig[A_\mu, A_\nu]$ with $T^l$ the $SU(N_c)$ generator 
in its fundamental representation and $\psi$ is a Dirac spinor with both color and flavor 
indices. The renormalized coupling constant is defined at the chemical 
potential via
\begin{equation}
g={24\pi^2\over (11N_c-2N_f\Big)\ln{\mu\over\Lambda}}
\label{eq:run}
\end{equation}
with $\mu>>\Lambda$ and $\Lambda=\Lambda_{\rm QCD}$ for $N_c=N_f=3$.

Perturbatively, the di-quark interaction is dominated by the process of one-
gluon exchange, as is shown in Fig. 1. 
The amplitude is simply that of the one-photon
exchange in QED multiplying the group theoretic factors 
$T_l^{c_1^\prime c_1}T_l^{c_2^\prime c_2}$, which can be decomposed into a color 
anti-symmetric channel (anti-triplet for $SU(3)$) and a color symmetric one 
(sextet for $SU(3)$), i.e.
\begin{equation}
T_l^{c_1^\prime c_1}T_l^{c_2^\prime c_2}
=-{N_c+1\over 4N_c}(\delta^{c_1^\prime c_1}\delta^{c_2^\prime c_2}
-\delta^{c_1^\prime c_2}\delta^{c_2^\prime c_1})
+{N_c-1\over 4N_c}(\delta^{c_1^\prime c_1}\delta^{c_2^\prime c_2}
+\delta^{c_1^\prime c_2}\delta^{c_2^\prime c_1})
\label{eq:decomp}
\end{equation}
As both the electric and magnetic parts of the one-photon exchange are 
repulsive between two electrons flying in opposite directions, the one-gluon 
exchange interaction between two quarks flying in opposite directions is attractive 
in the color anti-symmetric channel because of the negative sign of the first term 
on r. h. s. of (\ref{eq:decomp}). 

\begin{figure}[t]
\epsfxsize 5cm
\centerline{\epsffile{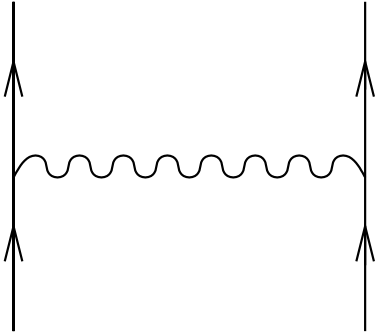}}
\medskip
\caption{The one-gluon exchange vertex.}
\label{One_gluon}
\end{figure}

At high baryon density, the Fermi sea of quarks tends to screen the one-gluon exchange 
interaction through the HDL(hard dense loop) resummed gluon propagator ~\cite{MLB}, which takes the 
form
\begin{equation}
D_{ij}^{ab}({\bf k},i\omega)={-i\delta^{ab}\over k^2+\omega^2+\sigma_M(k,\omega)}
\Big(\delta_{ij}-{k_ik_j\over k^2}\Big)
\label{eq:mag}
\end{equation}
\begin{equation}
D_{44}^{ab}({\bf k},i\omega)={-i\delta^{ab}\over k^2+\sigma_E(k,\omega)}
\label{eq:elec}
\end{equation}
and $D_{4j}({\bf k},i\omega)=0$ in Coulomb gauge with $i\omega$ the Matsubara energy of the 
gluon. In the region of energy and momentum for 
pairing, i.e. $k<<\mu$, $\omega<<\mu$ and $\omega<<k$, the magnetic self-energy is given by
\begin{equation}
\sigma_M(k,\omega)\simeq {\pi\over 4}m_D^2{|\omega|\over k}
\label{eq:magnetic}
\end{equation}
and the electric self-energy by
\begin{equation}
\sigma_E(k,\omega)\simeq m_D^2
\label{eq:electric}
\end{equation}
with $m_D^2={N_fg^2\mu^2\over 2\pi}$ the Debye mass. The absence of the screening of 
the long range magnetic gluon propagator in the static limit, $\omega=0$, leads to 
the forward singularity of the di-quark scattering, which  
dominates the pairing force. The long range magnetic gluons introduces also the non
Fermi liquid behavior, that renders the quark self energy function logarithmically enhanced 
toward the Fermi surface, i.e.
\begin{equation}
\Sigma(i\nu,p)\mid_{p=\mu}\simeq-\gamma_4{N_c^2-1\over N_c}
{g^2\over 24\pi^2}i\nu\ln{K_c^3\over \pi m_D^2
|\nu|}
\label{eq:nonFermi}
\end{equation} 
with $i\nu$ the Matsubara energy of the quark and $K_c$ ($k_BT<<K_c<<\mu$) a cutoff 
momentum. 

The energy scale of the color superconductivity is measured by the energy gap of the 
quark spectrum at $T=0$ or the superconducting transition temperature $T_c$. 
The determination of the latter quantity will be reviewed in this section. 

The proper vertex function $\Gamma(i\nu^\prime,{\bf p}^\prime |i\nu,{\bf p})$ for di-quark 
scattering with incoming energy-momenta $(p,\pm i\nu)$ and outgoing energy-momenta 
$(p^\prime,\pm i\nu^\prime)$ satisfies a Dyson-Schwinger equation, which, when projected 
into the color anti-symmetric channel with an angular momentum $J$, takes the form
\begin{equation}
\Gamma_J(i\nu^\prime,p^\prime |i\nu,p)=\tilde\Gamma_J(i\nu^\prime,p^\prime |i\nu,p)
+k_BT\sum_{\nu^{\prime\prime}}\int_0^\infty dq
K_J(i\nu^\prime,p^\prime |i\nu^{\prime\prime},q)
\Gamma_J(i\nu^{\prime\prime},q|i\nu,p),
\label{eq:DS}
\end{equation}
where $\tilde\Gamma_J(i\nu^\prime,p^\prime |i\nu,p)$ stands for the two particle
irreducible part of $\Gamma_J(i\nu^\prime,p^\prime |i\nu,p)$, the kernel
\begin{equation}
K_J(i\nu^\prime,p^\prime |i\nu^{\prime\prime},q)
={p^2\tilde\Gamma_J(i\nu^\prime,p^\prime |i\nu,p)\over 2\pi^2(2J+1)}S(i\nu,p)S(-i\nu,p)
\label{eq:kernel}
\end{equation}
with $S(i\nu,{\bf p})$ the full quark propagator. The spinor indices and their 
contraction structure have been suppressed in (\ref{eq:DS}) and 
(\ref{eq:kernel}). To the lowest order in coupling constant, 
$\tilde\Gamma$ is give by the one-gluon exchange diagram in 
Fig.1 and the quark propagators in the kernel (\ref{eq:kernel}) are replaced by bare ones. 
For $\nu<<\mu$, $\nu^\prime$, and $p\sim p^\prime\sim\mu$, we have
\begin{equation}
\Gamma_J(i\nu^\prime,p^\prime|i\nu,p)\simeq-{g^2\over 12\mu^2}\Big(1+{1\over N_c}\Big)
\Big(\ln{1\over|\hat\nu^\prime-\hat\nu|}+3c_J\Big)
\simeq-{g^2\over 12\mu^2}\Big(1+{1\over N_c}\Big)\Big(\ln{1\over|\hat\nu_>|}+3c_J\Big)
\label{eq:kernel_app}
\end{equation}
where
$\hat\nu={g^5\over 256\pi^4}\Big({N_f\over 2}\Big)^{5\over 2}{\nu\over\mu}$
with 
$c_J=1$ for $J=0$ and $c_J=\exp\Big(-2\sum_n^J{1\over n}\Big)$ for $J\neq 0$.
The last step of (\ref{eq:kernel_app}) follows from Son's approximation ~\cite{DTS}
with $\nu_>={\rm max}(\nu,\nu^\prime)$.

The pairing temperature within each angular momentum 
channel, $T_c^{(J)}$, corresponds 
to the highest temperature at which the Fredholm determinant of eq.(\ref{eq:DS}), 
$D_J={\rm det}(1-K_J)$ vanishes and the transition temperature is 
$T_c={\rm max}(T_c^{(J)})$ for all $J$. Here we have assumed the pairing between two quarks 
of the same herlicity, which contains the $s$-wave channel ($J=0$) and will take the maximum 
advantage of the pairing force.
  
To highlight the impact of the long range pairing force in the weak coupling
formula for $T_C$, we employ the standard expansion of the Fredholm 
determinant,
\begin{equation}
D_J=1-k_BT\sum_\nu\int_0^\infty dpK(i\nu,p|i\nu,p)+{(k_BT)^2\over 2}\sum_\nu\sum_{\nu^\prime}
\int_0^\infty dp\int_0^\infty dq\left|\,\matrix{K(i\nu,p|i\nu,p)&K(i\nu,p|i\nu^\prime,q)\cr
K(i\nu^\prime,q|i\nu,p)&K(i\nu^\prime,q|i\nu^\prime,q)\cr}\right|+...
\label{eq:fredholm}
\end{equation}
For pairing mediated by phonons, which is of a short range, the $m$-th term of the expansion is of the order 
of $g^{2m}\ln{\omega_D\over k_BT}$ with $\omega_D$ the Debye frequency. At the transition, 
$g^2\ln{\omega_D\over k_BT}\sim 1$ and the 
remaining terms are of the order of $g^{2m-2}$. This gives rise to the scaling formula 
$k_BT_c=c_{\rm BCS}\exp(-{\kappa_{\rm BCS}\over g^2})$ with $\kappa_{\rm BCS}$ fixed by the term of $m=1$ and the 
leading order of $c_{\rm BCS}$ fixed by the term of $m=2$. In case of QCD, the logarithm inside
the kernel (\ref{eq:kernel}) makes the $m$-th term of the expansion go like 
$g^{2m}\ln^{2m}{\mu\over k_BT}$ and the Fredholm determinant is dominated by a 
function of $g\ln{\mu\over k_BT}$. It follows then that the transition temperature scales 
like $k_BT_c=c_{\rm QCD}\exp\Big[-{\kappa_{\rm QCD}\over g}\Big]$ 
~\cite{DTS} ~\cite{RPDR1} ~\cite{DKH} leaving both $\kappa_{\rm QCD}$ and 
$c_{\rm QCD}$ non-trivial to determine.

A perturbative method was developed in ~\cite{BLR1} ~\cite{BLR2}, which yields the formula of 
the pairing temperature 
\begin{equation}
k_BT_c^{(J)}=cc^\prime c^{\prime\prime}c_J{\mu\over g^5}e^{-{\kappa\over g}}[1+O(g\ln g)],
\label{eq:Tc}
\end{equation}
where the exponent factor $\kappa=\sqrt{{6N_c\over N_c+1}}\pi^2$ was first obtained in ~\cite{DTS}, 
the pre-exponential factor $c=1024\sqrt{2}\pi^3N_f^{-{5\over 2}}$ was found in ~\cite{SW} and ~\cite{RPDR2}, 
the factor $c^\prime=2e^\gamma$ with $\gamma$ the Euler constant was found in ~\cite{RPDR2} and ~\cite{BLR2}. 
The factor $c^{\prime\prime}=\exp\Big[-{1\over 16}(\pi^2+4)(N_c-1)\Big]$ stems from the non 
Fermi liquid behavior of the quark self energy (\ref{eq:nonFermi}) and was calculated in ~\cite{BLR2} and 
reproduced in~\cite{WQDR} via the gap equation (The existence of this correction was suggested in ~\cite{DTS}). 
The gauge invariance of the formula (\ref{eq:Tc}) was discussed in ~\cite{BLR3}. The angular momentum dependent 
factor $c_J$ was calculated in ~\cite{BLR2} and the consequence of the universal exponent for 
different angular momentum channels in (\ref{eq:Tc}) was discussed in ~\cite{GLR} in the context of the  
crystalline superconductivity. Some higher order terms have also been identified ~\cite{CM}.
It follows from (\ref{eq:Tc}) that the superconducting transition temperature is $T_c=T_c^{(0)}$.

A similar perturbative method for the gap energy has been developed in ~\cite{TS}. 

\section{The Multiplicity of the Order Parameter.}
\label{sec:pert}
The Ginzburg-Landau theory is a powerful tool to explore the superconducting state right 
below the transition temperature. Unlike the ordinary superconductors, the order parameter of the 
color superconductivity which describes the di-quark condensate,  $\Psi_{f_1f_2}^{c_1c_2}$ is of 
multi-components, carrying the color-flavor indices 
and the chirality. For three color and three flavors, the symmetry group of theory in the chiral 
limit is $SU(3)_c\times SU(3)_{f_R}\times SU(3)_{f_L}
\times U(1)_B$ of which the electromagnetic gauge group, $U(1)_{\rm em}$ is a subgroup. 
Restricting within the even parity sector and 
neglecting the small projection in the color-sextet representation, the order parameter takes the form
\begin{equation}
{\Psi}^{c_1 c_2}_{f_1 f_2}
=\epsilon^{c_1 c_2 c}\epsilon_{f_1 f_2 f}\Phi^{c}_{f}.
\label{eq:order}
\end{equation}
and the Ginzburg-Landau free energy functional which is consistent with the symmetry group reads 
~\cite{KIGB} ~\cite{GR1}
\begin{equation}
\Gamma=\int d^3{\vec r}\Big[{1\over 4}F_{ij}^lF_{ij}^l
+{1\over 2}({\vec{\nabla}}\times{\vec{\cal A}})^2+
4{\rm{tr}}({\vec D}\Phi)^{\dag}({\vec D}\Phi)
+4a{\rm{tr}}\Phi^{\dag}\Phi+b_1{\rm{tr}}(\Phi^{\dag}\Phi)^2 
+b_2({\rm{tr}}\Phi^{\dag}\Phi)^2\Big]
\label{eq:free}
\end{equation}
where the covariant derivative ${\vec D}\Phi={\vec\nabla}\Phi-ig{\vec A}
\Phi+i{2\over {\sqrt 3}}e{\vec{\cal A}}\Phi T^8$ with $\vec A^l$ the color vector potential and 
$\vec {\cal A}$ the ordinary electromagnetic vector potential. The $3\times 3$ matrix $\Phi=
\Phi_0+\Phi_lT^l$ with $\Phi_0$ the singlet and $\Phi_l$ the octet under a simultaneous 
color-flavor rotation. The generator here, $T^l$, should be understood to be that of the 
anti-triplet representation of $SU(3)_c$. At weak coupling, the one-gluon 
exchange lends us the following expressions of the Ginzburg-Landau 
coefficients ~\cite{KIGB} ~\cite{GR1}
\begin{eqnarray}
a&={48\pi^2\over 7\zeta(3)}k_B^2T_C(T-T_C),\nonumber \\
b_1&=b_2={576\pi^4\over 7\zeta(3)}\Big({k_BT_C\over\mu}\Big)^2,
\label{eq:coeff}
\end{eqnarray}
which follows also from the mean field approximation of the NJL effective 
action at moderate coupling if $k_BT_c<<\mu$. 

Let us review first the structure of a homogeneous condensate ~\cite{KIGB}, i.e. 
$\vec A=\vec {\cal A}=\vec\nabla\Phi=0$, for which three regions of the parameter 
space need to be considered respectively. 1) $b_1>0$ and $b_1+3b_2>0$
The minimum free energy corresponds to the
color-flavor locked condensate ~\cite{ARW} with $\Phi=\phi_0U$, where
\begin{equation}
\phi_0=\sqrt{-{2a\over b_1+3b_2}},
\label{eq:cfl}
\end{equation} 
and $U$ is a unitary matrix (The nontrivial winding of $U$ onto the gauge group, 
$SU(3)_c\times U(1)_{\rm em.}$ gives rise to vortex filaments ~\cite{GR1}). 
2) $b_1<0$ but $b_1+b_2>0$: In this case the 
color-flavor locked condensate
(\ref{eq:cfl}) becomes a saddle point of the free energy, which
is nevertheless bounded from below. The minimum corresponds to an
isoscalar condensate ~\cite{KIGB}, given by
$\Phi={\rm diag}\Big(\sqrt{-{2a\over b_1+b_2}}e^{i\alpha},0,0\Big)$.
3) For $b_1$ and $b_2$ outside the region specified by 1) and 2), the free 
energy is no longer bounded from below.
Higher powers of the order parameter have
to be restored and the superconducting transition becomes the first order one.
In the following we shall focus our attention to the case 1) of the parameters.

The color-flavor locked condensate breaks the original symmetry group down to 
a $SU(3)$ of simultaneous color-flavor rotation, among which there is an 
unbroken $U(1)$ gauge symmetry and the corresponding gauge potential $\vec{\cal V}$ 
is obtained through the rotation ~\cite{ABR}
\begin{eqnarray}
{\vec{\cal V}}&=-{\vec A}^8{\sin\theta}+{\vec{\cal A}}{\cos{\theta}}\nonumber \\
{\vec V}&={\vec A}^8{\cos\theta}+{\vec{\cal A}}{\sin{\theta}},
\label{eq:elecwk}
\end{eqnarray}
with $\tan\theta=-{{2e}\over {{\sqrt 3}g}}$. In comparison with the electroweak theory, 
the field $\vec{\cal V}$ and $\vec V$ are the analogs of the photon and the $Z$-boson, 
and the rest of the $A$'s are like the $W$-boson's. The mass of the $\vec V$ 
and that of $\vec A^l$ $(l=1,...,7)$ are  
$$
m_Z^{2}=4g^2{\phi_0^2}{\rm sec}^2\theta={1\over {\delta}^2}, \qquad
m_W^2=m_Z^{2}{\cos^2{\theta}}={1\over {\delta^\prime}^2}.
\label{eq:vector}
$$
with $\delta$ and $\delta^\prime$ the corresponding penetration depths. Other low-lying 
excitations of even parity consist of a Goldstone boson associated to the broken baryon 
number, $U(1)_B$, the 
Higgs bosons associated to $\vec V$ and to $\vec A^l$ $(l=1,...,7)$ with masses
$$
m_H^2=(b_1+3b_2)\phi_0^2={2\over {\xi}^2}, \qquad 
m_H^{\prime 2}=b_1\phi_0^2={2\over {\xi^\prime}^2},
\label{eq:higgs}
$$
and ($\xi$, $\xi^\prime$) the corresponding coherence lengths.
The excitations of odd parities includes the Goldstone bosons associated to 
the chiral symmetry breaking triggered by the color-flavor locking ~\cite{ARW}.

An interesting issue to address is the type of the superconductivity in response to an external 
magnetic field, which is partially screened because of its projection onto the broken $\vec V$
field through (\ref{eq:elecwk}). This amounts to calculate the free energy of a domain wall 
separating the super phase and the normal phase with the bulk of super phase and that of the normal 
phase held in thermal equilibrium under an external magnetic field of the critical strength,
\begin{equation}
H_c=2\sqrt{6a^2\over b_1+3b_2}|{\rm csc}\theta|.
\label{eq:criticalH}
\end{equation}
The simplest ansatz of the solution to the 
Ginzburg-Landau equations that minimizes the domain wall free energy consists of only 
nonzero $\vec V$, parallel to the external magnetic field, and nonzero components $\Phi_0$ and $\Phi_8$, 
which maintains the maximum symmetry spared by the boundary condition. Upon introducing 
dimensionless quantities via
$$
s={\delta}x, \quad \Phi_0+{1\over {\sqrt 3}}\Phi_8={\phi}_0u,
\quad \chi={\sqrt 2}(\Phi_0-{1\over {2\sqrt 3}}\Phi_8)={\sqrt 2}{\phi}_0v, \quad
{V}=-{\sqrt{-3a}\over g}A{\cos\theta}.
\label{eq:rewe}
$$
the corresponding Ginzburg-Landau equations become
\begin{eqnarray}
&-A^{\prime\prime}+{1\over 3}A(2u^2+v^2)=0, \nonumber \\
&-{1\over {\kappa}^2}u^{\prime\prime}+(A^2-1)u
+{1\over 3}(2u^2+v^2)u+{1\over 3}
\rho(u^2-v^2)u=0, \nonumber \\
&-{1\over {\kappa}^2}v^{\prime\prime}
+{1\over 4}(A^2-4)v+{1\over 6}(u^2+5v^2)v
-{1\over 6}\rho(u^2-v^2)v=0,
\label{eq:frion}
\end{eqnarray}
subject to the boundary conditions
\begin{eqnarray}
&u \mapsto 0, \quad v \mapsto 0, \quad A^{\prime} \mapsto 1 \quad
{\rm at} \quad s \mapsto -\infty \nonumber \\
&u \mapsto 1, \quad v \mapsto 1
, \quad A^{\prime} \mapsto 0 \quad
{\rm at} \quad s, \mapsto \infty
\label{eq:erionb}
\end{eqnarray}
where $\kappa={\delta\over\xi}$ is the Ginzburg-Landau parameter, 
$\rho={{b_1-3b_2}\over {b_1+3b_2}}$ with $\rho=-{1\over 2}$ for mean field approximation,
and the prime denotes the derivative with respect to $s$.
In case of an ordinary superconductors with one component of the order 
parameter, Ginzburg-Landau found analytically ~\cite{GL} that the domain wall energy vanishes at
\begin{equation}
\kappa=\kappa_c={1\over\sqrt{2}}\simeq 0.707,
\label{eq:GLparam}
\end{equation}
which was later clarified by Arikosov ~\cite{ABRI} as the demarcation between the type I 
superconductivity ($\kappa<\kappa_c$) and the type II one ($\kappa>\kappa_c)$. 
The equations (\ref{eq:frion}) are more complicated and depend on two dimensionless 
parameters, $\kappa$ and $\rho$. Nevertheless a set of inequalities have been established 
following the variational 
arguments ~\cite{GR2}. We have shown that the domain wall energy is a decreasing function of 
$\kappa$ and an increasing function of $\rho$. Consequently,
$$
\kappa_c(\rho)\le {1\over\sqrt{2}}, \quad {d\kappa_c\over d\rho}\ge 0.
\label{eq:theorem}
$$
The numerical solution of the equations (\ref{eq:erionb}) shows that
\begin{equation}
\kappa_c=0.589
\label{eq:number}
\end{equation}
for $\rho=-{1\over 2}$. While the color-flavor locked component $\Phi_0$ dominates the bulk 
of the condensate, the unlocked component, $\Phi_8$, shows up near the domain wall.

Various fluctuation effects of the Ginzburg-Landau theory have also been considered in
the literature, see e.g. ~\cite{GR2} ~\cite{New}.

\section{Concluding Remarks}
In this lecture, I have reviewed some novel properties of the color 
superconductivity. A systematical approach to determine various physical quantities 
at weak coupling has been sketched. While the accuracy 
of the perturbative formula for 
the transition temperature (\ref{eq:Tc}) and that for the Ginzburg-Landau 
coefficients (\ref{eq:coeff}) is uncertain when extrapolated to a realistic 
quark matter, it is instructive to substitute in the observationally 
attainable quark density for an order of magnitude estimation of the 
quantities of interests. 

Taking the generic chemical potential of in the core of a neutron star, $\mu=400$MeV, 
as a bench mark, we have $\alpha_S\equiv {g^2\over 4\pi}\simeq 1$ and the 
mixing angle $\theta=-5.6^\circ$ for $N_c
=N_f=3$ and $\Lambda=\Lambda_{\rm QCD}=200$MeV. Using eq. (\ref{eq:coeff}) and 
(\ref{eq:criticalH}), we find the 
thermodynamical critical field $H_c=1.47\times 10^{20}\Big({k_BT_c\over\mu}
\Big)\Big(1-{T\over T_c}\Big)$Gauss. Furthermore, the criterion 
(\ref{eq:number}) implies that the color superconductivity is type I if 
$k_BT_C<14$MeV and type II if $k_BT_c>14$MeV. It follows from (\ref{eq:Tc}) 
that $k_BT_c\simeq 3.5$MeV with one-gluon exchange approximation and the color 
superconductivity is of type I. Beyond the one-gluon exchange approximation, 
$k_BT_c$ could be high enough to cross over to the type II region, but it is 
unlikely to be a strong type II system. The critical magnetic field, 
which is by orders of magnitude higher than the typical 
magnetic field in a neutron star, is too strong to lead to observational 
distinctions between the two types of the color-superconductivity.

\section{Acknowledgments}
I am grateful to W. Brown, I. Giannakis and J. T. Liu for collaborations. This 
work is supported in part by US Department of Energy contract number DE-FG02
-91ER40651-TASKB.

\ifpreprintsty\else
\fi

\end{document}